\documentclass[final,5p,times,twocolumn]{elsarticle}

\usepackage{graphicx}

\usepackage{amssymb}

\journal{Nuclear Instruments and Methods in Physics Research, A}

\begin{document}

\begin{frontmatter}



\title{The LOPES experiment --- recent results, status and perspectives}


\author[1]{T.~Huege}
\author[1]{W.D.~Apel}
\author[2,14]{J.C.~Arteaga}
\author[3]{T.~Asch}
\author[4]{L.~B\"ahren}
\author[1]{K.~Bekk}
\author[5]{M.~Bertaina}
\author[6]{P.L.~Biermann}
\author[1,2]{J.~Bl\"umer}
\author[1]{H.~Bozdog}
\author[7]{I.M.~Brancus}
\author[8]{P.~Buchholz}
\author[4]{S.~Buitink}
\author[5,9]{E.~Cantoni}
\author[5]{A.~Chiavassa}
\author[1]{K.~Daumiller}
\author[2,15]{V.~de Souza}
\author[1]{P.~Doll} 
\author[1]{R.~Engel}
\author[4,10]{H.~Falcke} 
\author[1]{M.~Finger} 
\author[11]{D.~Fuhrmann}
\author[3]{H.~Gemmeke}
\author[8]{C.~Grupen}
\author[1]{A.~Haungs}
\author[1]{D.~Heck}
\author[4]{J.R.~H\"orandel}
\author[4]{A.~Horneffer}
\author[2]{D.~Huber}
\author[1,16]{P.G.~Isar}
\author[11]{K.-H.~Kampert}
\author[2]{D.~Kang}
\author[3]{O.~Kr\"omer}
\author[4]{J.~Kuijpers}
\author[4]{S.~Lafebre}
\author[2]{K.~Link}
\author[12]{P.~{\L}uczak}
\author[2]{M.~Ludwig}
\author[1]{H.J.~Mathes}
\author[2]{M.~Melissas}
\author[9]{C.~Morello}
\author[1]{S.~Nehls}
\author[1]{J.~Oehlschl\"ager}
\author[2]{N.~Palmieri}
\author[1]{T.~Pierog}
\author[11]{J.~Rautenberg}
\author[1]{H.~Rebel}
\author[1]{M.~Roth}
\author[3]{C.~R\"uhle}
\author[7]{A.~Saftoiu}
\author[1]{H.~Schieler}
\author[3]{A.~Schmidt}
\author[1]{F.G.~Schr\"oder}
\author[13]{O.~Sima}
\author[7]{G.~Toma}
\author[9]{G.C.~Trinchero}
\author[1]{A.~Weindl}
\author[1]{J.~Wochele}
\author[1]{M.~Wommer}
\author[12]{J.~Zabierowski}
\author[6]{J.A.~Zensus}

\cortext[cor]{Corresponding author: Tim Huege $<$tim.huege@kit.edu$>$}

\address[1]{Karlsruhe Institute of Technology (KIT) - Campus North, Institut f\"ur Kernphysik, Germany}
\address[2]{Karlsruhe Institute of Technology (KIT) - Campus South, Institut f\"ur Experimentelle Kernphysik, Germany}
\address[3]{Karlsruhe Institute of Technology (KIT) - Campus North, Institut f\"ur Prozessdatenverarbeitung und Elektronik, Germany}
\address[4]{Radboud University Nijmegen, Department of Astrophysics, The Netherlands}
\address[5]{Dipartimento di Fisica Generale dell' Universita Torino, Italy}
\address[6]{Max-Planck-Institut f\"ur Radioastronomie Bonn, Germany}
\address[7]{National Institute of Physics and Nuclear Engineering, Bucharest, Romania}
\address[8]{Universit\"at Siegen, Fachbereich Physik, Germany}
\address[9]{INAF Torino, Istituto di Fisica dello Spazio Interplanetario, Italy}
\address[10]{ASTRON, Dwingeloo, The Netherlands}
\address[11]{Universit\"at Wuppertal, Fachbereich Physik, Germany}
\address[12]{Soltan Institute for Nuclear Studies, Lodz, Poland}
\address[13]{University of Bucharest, Department of Physics, Bucharest, Romania}

\address[14]{\scriptsize{now at: Universidad Michoacana, Instituto de F\'{\i}sica y Matem\'aticas, Mexico}}
\address[15]{\scriptsize{now at: Universidade S$\tilde{a}$o Paulo, Instituto de F\'{\i}sica de S\~ao Carlos, Brasil}}
\address[16]{\scriptsize{now at: Institute of Space Science, Bucharest, Romania}}

\begin{abstract}
The LOPES experiment at the Karlsruhe Institute of Technology has been taking radio data in the frequency range from 40 to 80 MHz in coincidence with the KASCADE-Grande air shower detector since 2003. Various experimental configurations have been employed to study aspects such as the energy scaling, geomagnetic dependence, lateral distribution, and polarization of the radio emission from cosmic rays. The high quality per-event air shower information provided by KASCADE-Grande has been the key to many of these studies and has even allowed us to perform detailed per-event comparisons with simulations of the radio emission. In this article, we give an overview of results obtained by LOPES, and present the status and perspectives of the ever-evolving experiment.
\end{abstract}


\end{frontmatter}


\section{Introduction}

Forty years after the initial discovery of radio emission from cosmic ray air showers \citep{JelleyFruinPorter1965}, cosmic ray radio detection has once again become a very active field of research. The LOPES experiment \citep{FalckeNature2005} in particular has revived the radio detection activities with an innovative approach combining digital radio-interferometry with detailed air shower measurements employing a classical particle detector array. The experiment is situated at Campus North of the Karlsruhe Institute of Technology, at the site of the KASCADE-Grande air shower experiment \citep{AntoniApelBadea2003,ApelArteagaBadea2010}. The close integration of LOPES with KASCADE-Grande, leading in particular to the availability of high-quality per-event air shower parameters, has proven to be the key to many of the successes of LOPES. In comparison, the CODALEMA experiment \citep{ArdouinBelletoileCharrier2005} in Nan\c{c}ay is situated at an observing site with a much quieter radio background, facilitating the detection of radio signals in individual antennas without resorting to interferometric techniques. The CODALEMA experiment, however, does not have access to per-event air shower information of as high a quality as LOPES.

In this article, we first describe the basic properties of the LOPES experiment, followed by an overview of the analysis procedures we apply to our data before we finally discuss the different phases and results of the ever-evolving experiment.


\section{Basic properties}

LOPES is a digital radio-interferometer measuring in the 40--80~MHz frequency window. This frequency window was chosen to avoid FM radio-transmitters at frequencies above 80~MHz and short-wave and atmospheric noise at frequencies below 40~MHz. All LOPES channels are continuously sampled with 12 bit ADCs and a sampling rate of 80 MHz, i.e., in the second Nyquist zone. The data are stored in ring buffers, and when a trigger arrives from the KASCADE or Grande arrays, 0.8~ms of data are read out and stored for each channel. Events with energies above $\approx 10^{16}$~eV are triggered; the threshold for radio detection lies significantly higher. While these basic properties are universal to all configurations of LOPES, the layout and type of antennas has changed over the course of time.

\begin{figure*}[ht]
  \centering
  \includegraphics[width=.8\textwidth]{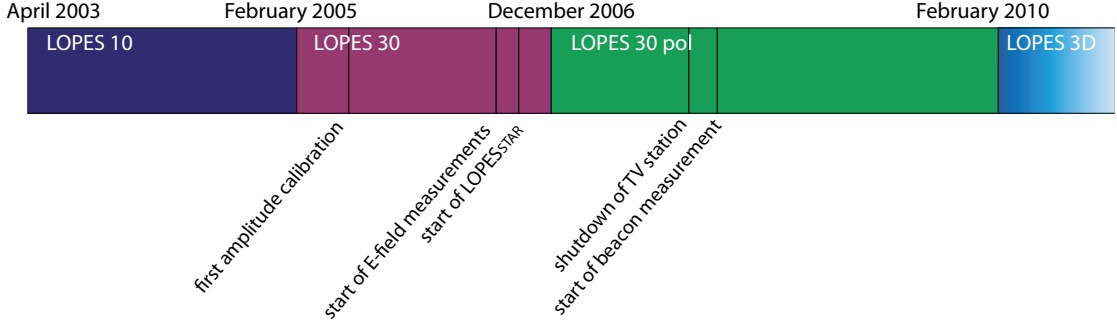}
  \caption{Time-line of the evolution of the LOPES experiment. LOPES$^{\mathrm{STAR}}$ commenced in late 2005 and runs in parallel with the other phases of LOPES.}
  \label{fig:timeline}
\end{figure*}

In Fig.\ \ref{fig:timeline}, an overview is given of the different phases of LOPES: LOPES~10, LOPES~30, LOPES~30~pol and LOPES~3D. The different phases were tailored to different scientific questions, and will be discussed in some more detail in the following sections.

\begin{figure}[htb]
  \centering
  \includegraphics[width=.45\textwidth]{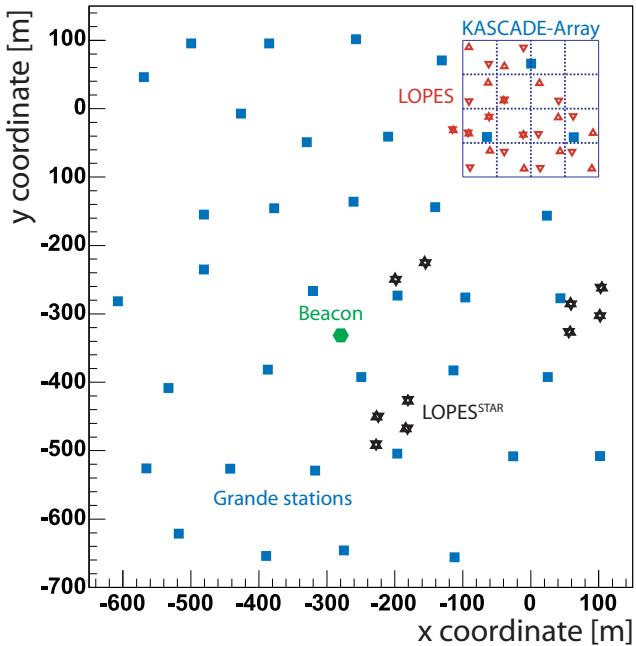}
  \caption{Layout of the LOPES~30 pol setup. Triangles denote east-west- and north-south-aligned antennas, respectively; stars denote positions with both an east-west and north-south-aligned antenna.}
  \label{fig:lopeslayout}
\end{figure}

Although the layout of the antenna array has been changing over the different phases, in all cases the antennas have been placed in a region coinciding with the KASCADE array of KASCADE-Grande, as depicted for the LOPES~30~pol setup in Fig.\ \ref{fig:lopeslayout}. The antennas used for all phases except LOPES~3D have been inverted-V dipole antennas based, like most of the electronics used, on prototype designs for LOFAR \citep{FalckevanHaarlemdeBruyn2007}. Further technical details on the hardware of LOPES can be found in reference \citep{HornefferThesis2006}.

\section{Analysis procedure} \label{sec:analysisprocedure}

The analysis procedure of LOPES data exhibits a high degree of sophistication, which is necessary in particular because of the comparatively high radio-frequency interference background at the ``industrial'' location of the KIT campus north. Here, we shortly discuss the analysis steps applied in our standard reconstruction.

\begin{enumerate}
\item{An absolute amplitude calibration is applied to the data. As of LOPES~30, this is based on a measurement with an external, calibrated reference source \citep{NehlsHakenjosArts2007}. The remaining systematic uncertainty of electric field amplitudes between events is of order 10~\%.}
\item{To suppress narrow-band transmitters, a digital filtering algorithm suppressing peaks in the frequency-spectra of each antenna is applied. This hardly affects the broad-band cosmic ray radio signals, and therefore increases the signal-to-noise ratio significantly.}
\item{The delay of each individual channel (dominated by cable delays and determined in a dedicated calibration campaign) is corrected for.}
\item{A determination of the relative phases of the beacon transmitter operated in the vicinity of the LOPES antennas (cf.\ Fig.\ \ref{fig:lopeslayout}) is used to constrain the relative timing of the LOPES channels to $\approx 1$~ns precision \citep{SchroederAschBaehren2010}. Such a high precision is a necessary prerequisite for a reliable interferometric analysis. Before the availability of the beacon transmitter, a TV transmitter in the LOPES band was used for the phase calibration.}
\item{Dispersion introduced by the instrumental response (in particular the bandpass filters) is de-convoluted from the signal traces, increasing the signal-to-noise ratio of cosmic ray radio pulses.}
\item{The 12.5~ns sampled raw data are up-sampled (i.e., interpolated correctly) to a higher sampling rate. This is possible because the data were sampled correctly in the second Nyquist zone and therefore contain the full information in the 40--80~MHz frequency window.}
\item{A digital beam-forming is applied to arrange the time-series data of all channels correctly for radio emission received from the presumed arrival direction. (For the first iteration, the direction reconstructed by KASCADE or Grande is used as a starting point. Likewise, the core position is taken from the KASCADE or Grande reconstruction.)}
\item{The antenna characteristics (i.e., frequency-dependent gain) for the established arrival direction is corrected for.}
\item{The cross-correlation beam is calculated using eq. (\ref{eqn:ccbeam}). Afterwards, it is block-averaged over 37.5~ns (approximately the width of the LOPES impulse response) and fitted with a Gaussian. The height of the Gaussian is used as a measure for the amplitude of the overall coherent radio signal.}
\item{For high signal-to-noise events, the maximum electric field amplitude in each antenna in a time-window around the CC-beam maximum is determined for the study of per-event lateral distributions.}
\item{A correlation with a database listing events recorded during thunderstorms is performed.}
\end{enumerate}

Steps 7-9 are performed in an iterative loop, during which the arrival direction of the emission and the spherical curvature of the electromagnetic front are varied to maximize the value of the CC-beam. Typical values for the spherical curvature of the electromagnetic wavefront are in the range from $\approx 2$--10~km \citep{FalckeNature2005}. A plane-wave assumption produces significantly poorer coherence.
The CC-beam for each sample is calculated as
\begin{equation} \label{eqn:ccbeam}
cc[t] = \, ^+_- \sqrt{\left|\frac{1}{N_{\mathrm{pairs}}}\sum^{N-1}_{i=1}\sum^{N}_{j>i}s_i
[t]s_j[t]\right|}
\end{equation}
with $N$ the number of antennas, $s_{n}[t]$ the time shifted (cf.\ step 6 above) data of the denoted antenna, and $N_{\mathrm{pairs}}$ the number of unique pairs of antennas. The sign for each sample is the sign it had before taking the square root.
\begin{figure*}[ht]
  \centering
  \includegraphics[width=.45\textwidth]{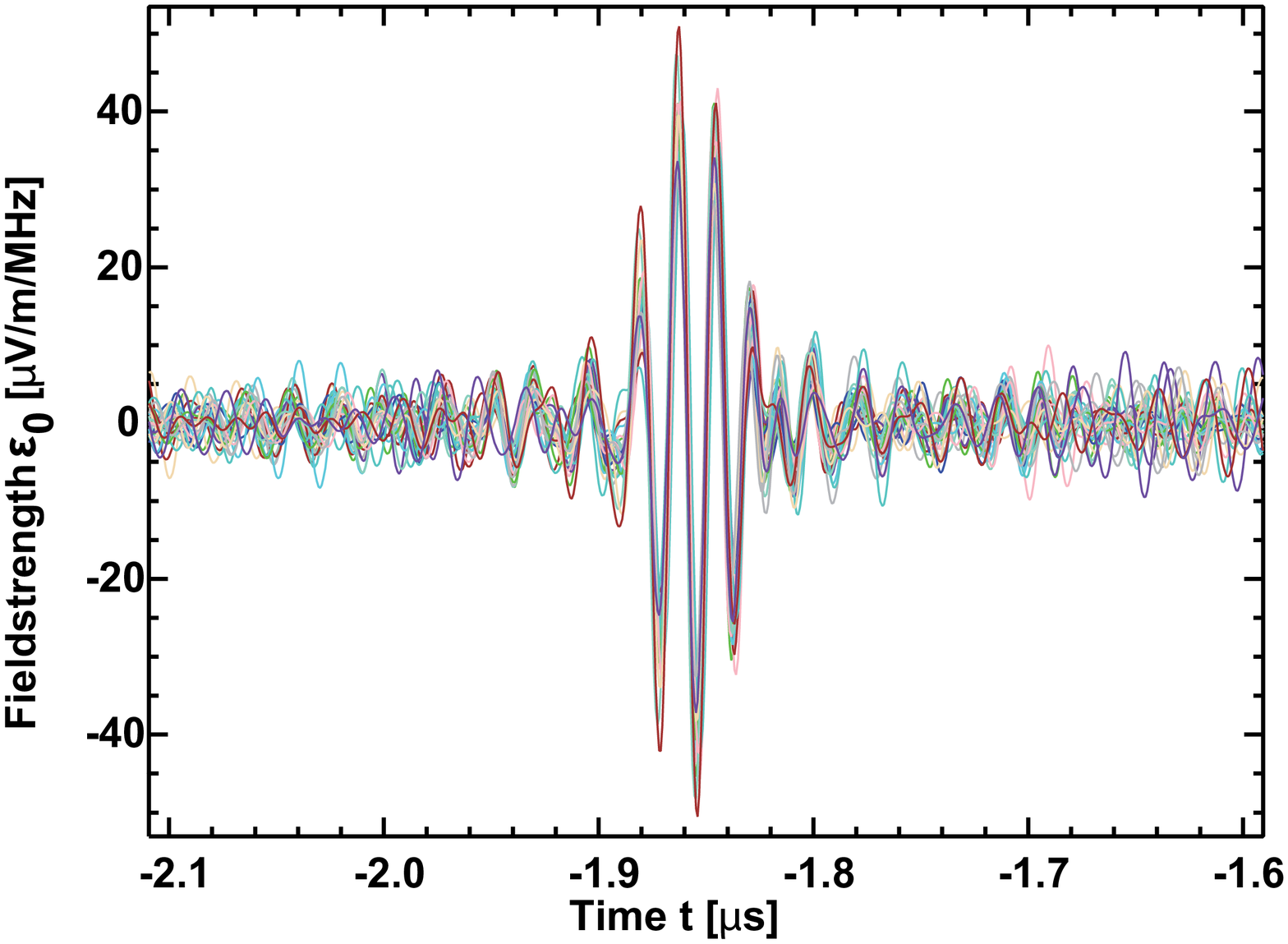}
  \includegraphics[width=.45\textwidth]{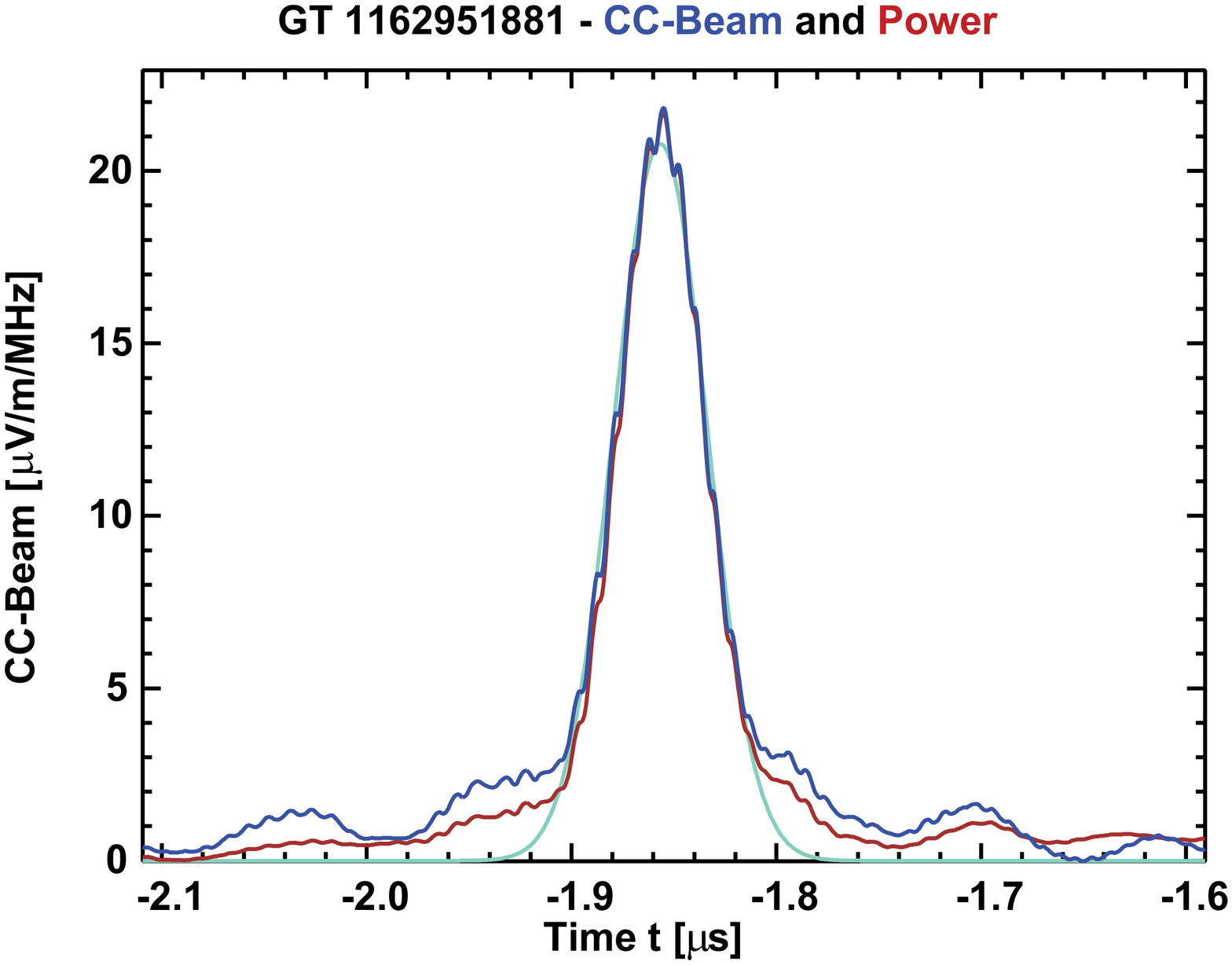}
  \caption{Left: Up-sampled and beam-formed cosmic ray radio pulses seen in individual LOPES antennas. Right: Block-averaged cross-correlation-beam calculated from the individual antenna signals (blue) with its Gauss fit (light blue). The brown line denotes the power-beam, a quantity describing the total power received in all antennas together. As the cross-correlation-beam and power-beam coincide, the power in the radio signal is fully coherent.}
  \label{fig:ccbeam}
\end{figure*}


\section{LOPES~10}

\begin{figure}[htb]
  \centering
  \includegraphics[width=.31\textwidth,angle=270,clip=true]{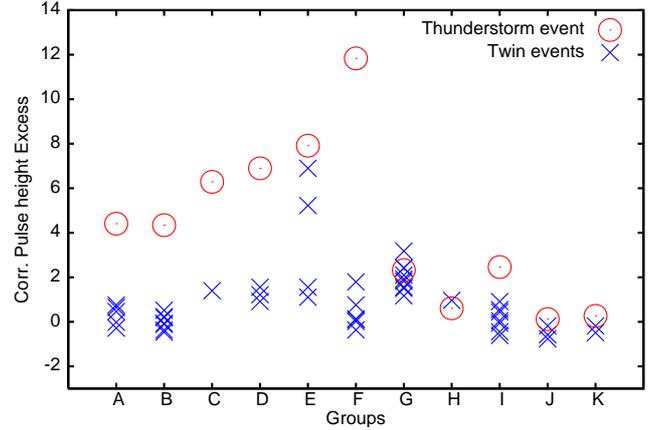}
  \caption{Comparison of events measured during thunderstorms (red circles) with a set of ``twin'' events with similar parameters taken during fair weather conditions (blue crosses). The $y$-axis quantifies the deviation of the measured field strength from the field strength expected for an event with the given parameters. During thunderstorms, LOPES clearly sees enhanced radio emission from cosmic ray air showers.}
  \label{fig:thunderstorms}
\end{figure}

The initial phase of LOPES consisted of 10 inverted-V dipole antennas oriented in the east-west direction. Only a relative amplitude calibration was available for these channels. With LOPES~10, the proof-of-principle for radio detection with a digital radio-interferometer was made \citep{FalckeNature2005}. Additional results were a near-linear correlation of the radio field strength with the primary particle energy, confirming the expected coherence of the radio emission, a clear correlation of the field strength with the angle between magnetic field and shower axis (the so-called ``geomagnetic angle'', both results published in \citep{FalckeNature2005}), and an approximately exponential damping of the radio emission with lateral distance with a slope parameter of $R_{0} = (230\pm51)$~m \citep{ApelAschBadea2006}. In the latter publication, it was also demonstrated that radio signals could be detected up to lateral distances of $>500$~m and that as of $10^{17}$~eV, LOPES has a high detection efficiency. Furthermore, it was demonstrated with LOPES~10 that very highly inclined air showers with zenith angles up to above 80$^{\circ}$ can be well detected with radio antennas \citep{PetrovicApelAsch2006} and that the radio emission from air showers can be significantly enhanced during thunderstorms, see Fig.\ \ref{fig:thunderstorms} and \citep{BuitinkApelAsch2006}. In contrast, fair weather atmospheric electric fields do not significantly affect the radio emission.


\section{LOPES~30}

After the success of LOPES~10, the experiment was extended to 30 inverted-V dipole antennas aligned in the east-west direction. The higher number of antennas distributed over a larger area increased the sensitivity (number of antennas) and angular resolution (longest baseline) of the array. In addition, the new layout was aimed at performing detailed per-event lateral-distribution studies. The reconfiguration of LOPES was complemented with regular calibration campaigns using an external reference source, thereby providing an unprecedented absolute calibration of each individual detection channel \citep{NehlsHakenjosArts2007} with a remaining systematic uncertainty between events of $\approx 10$\% with respect to the electric field strength. To further improve the understanding of the instrumental response and possible environmental effects on the radio emission, a thorough monitoring of all relevant environmental parameters was started. This included in particular a monitoring of the atmospheric electric field to be able to reliably identify events recorded during thunderstorms.

Results from LOPES~30 are plentiful, and we only present a few highlights here. The correlations of the radio electric field strength with primary particle energy, lateral distance and geomagnetic angle were studied once more using the absolute calibrated, east-west polarization LOPES~30 data \citep{HornefferICRC2007}. The results are depicted in Fig.\ \ref{fig:lopes30iterative} and can be parameterized with the following formula:
\begin{figure}[h!tb]
  \centering
  \includegraphics[width=.29\textwidth,angle=270]{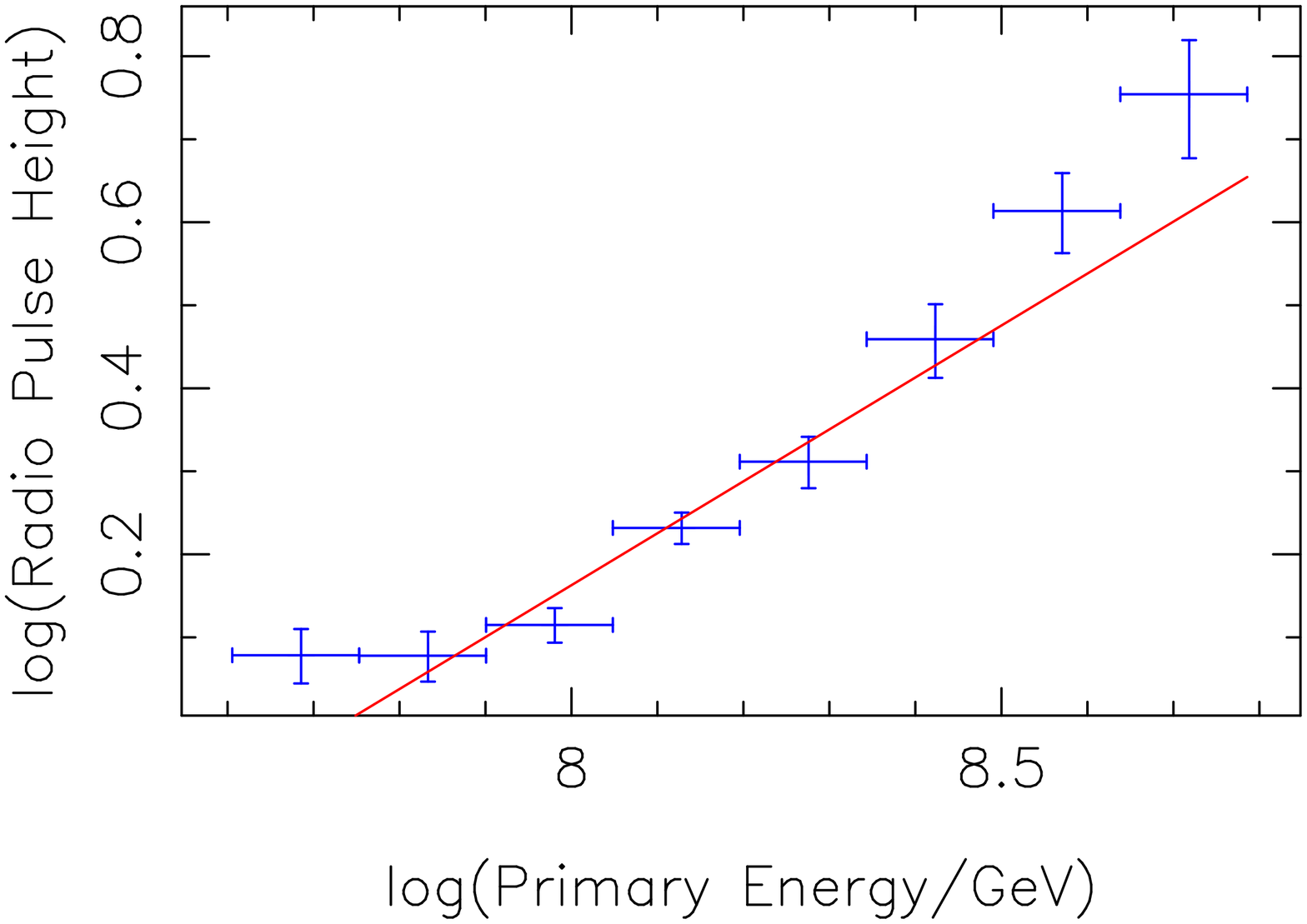}\vspace{0.3cm}
  \includegraphics[width=.29\textwidth,angle=270]{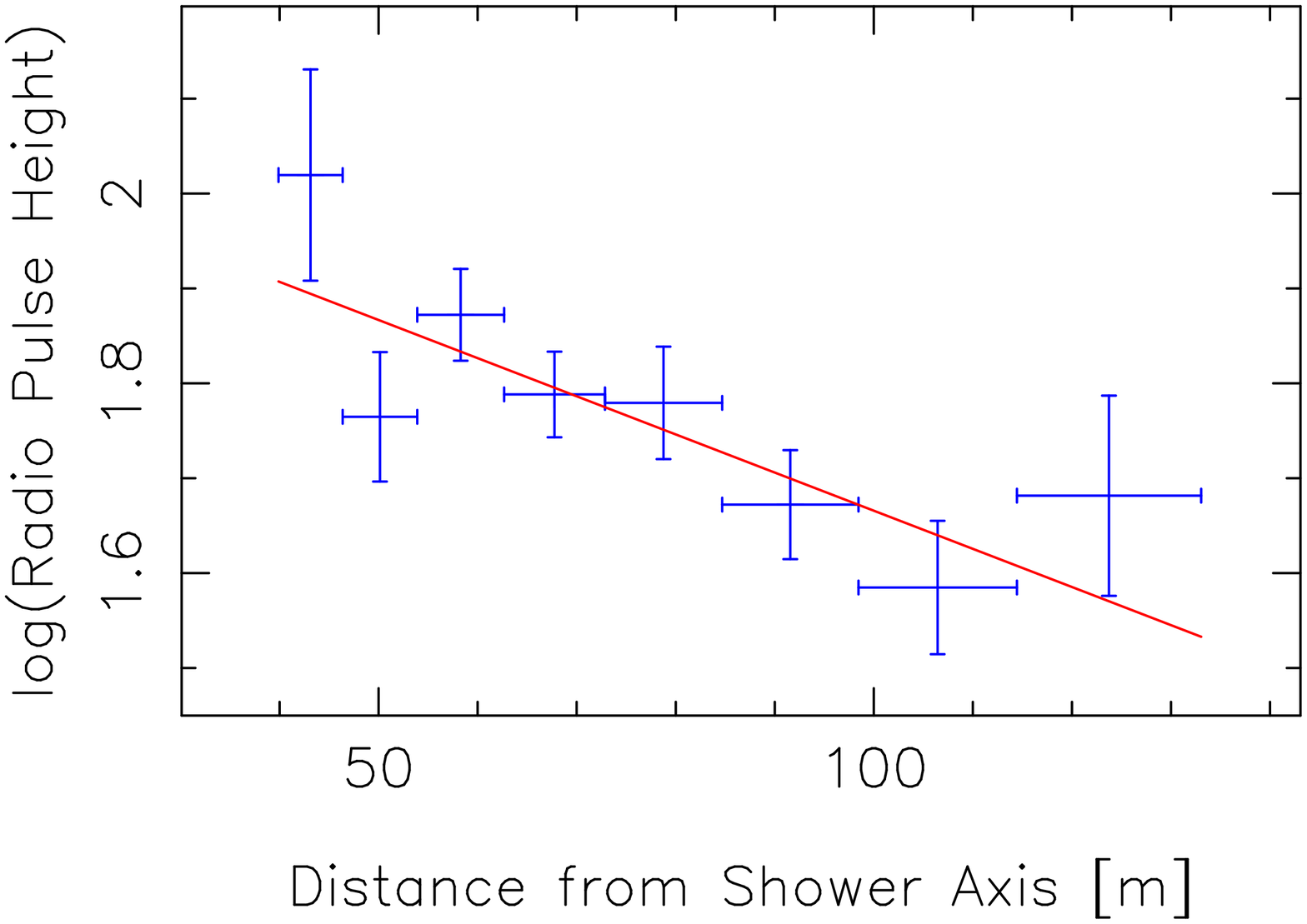}\vspace{0.3cm}
  \includegraphics[width=.29\textwidth,angle=270]{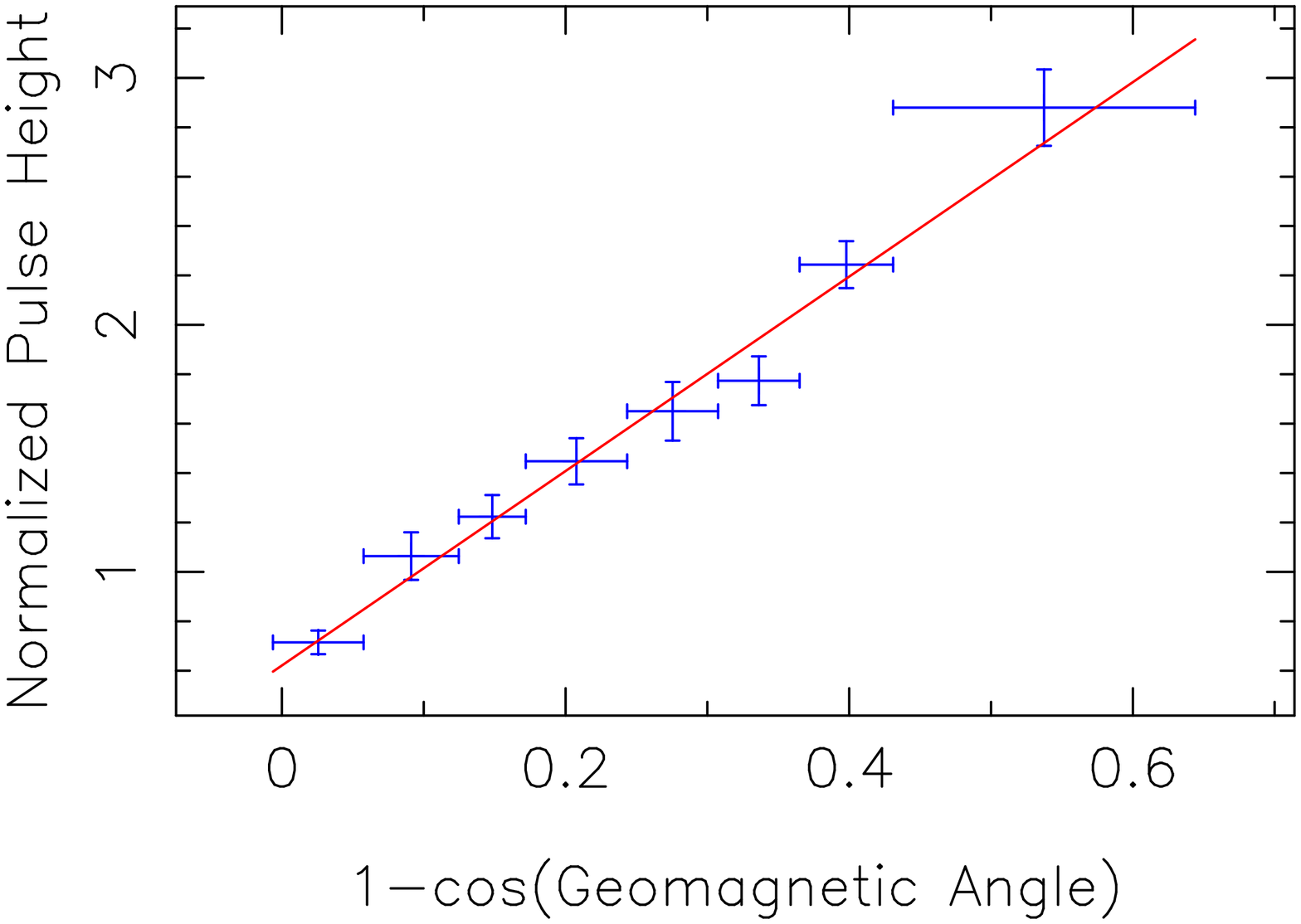}
  \caption{Correlations of the LOPES~30 east-west polarization electric field strengths with primary energy (left), lateral distance (middle) and geomagnetic angle (right).}
  \label{fig:lopes30iterative}
\end{figure}
\begin{eqnarray}
&\epsilon_{\rm est}  =
(11\pm1.)
\left((1.16\pm0.025)-\cos\alpha\right) \cos\theta & \nonumber \\
& \exp\left(\frac{\rm -R_{SA}}{\rm (236\pm81)\,m}\right)
\left(\frac{\rm E_{p}}{\rm 10^{17}eV}\right)^{(0.95\pm0.04)} 
\left[\frac{\rm \mu V}{\rm m\,MHz}\right]&
\label{eq:horneffer-energy}
\end{eqnarray}
with $\alpha$ the geomagnetic angle, $\theta$ the zenith angle, ${\rm R_{SA}}$ the mean distance of the antennas to the shower axis, and ${\rm E_{p}}$ the primary particle energy. The given errors are the statistical errors from the fit.

Another study dealt with the determination of the direction resolution, which corresponds to $\approx 1.3^{\circ}$ and is probably limited by our current understanding of the shape of the electromagnetic radio wave front \citep{NiglApelArteaga2008a}. Frequency spectra of cosmic ray radio signals were investigated with LOPES~30 as well \citep{NiglApelArteaga2008b}, which confirmed the expectation that the signal falls off to higher frequencies.

\begin{figure*}[htb]
  \centering
  \includegraphics[width=.45\textwidth]{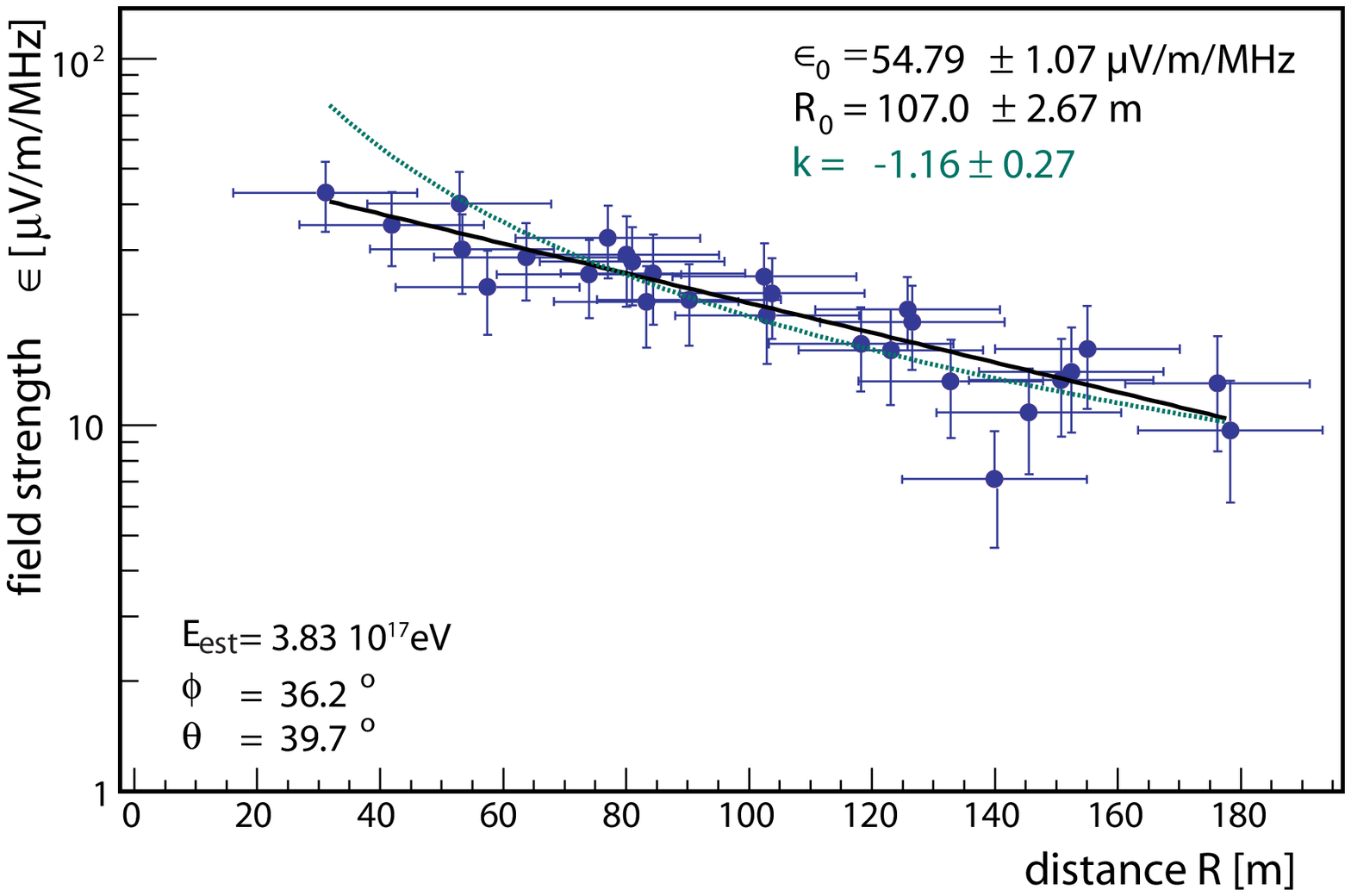}\hspace{0.5cm}
  \includegraphics[width=.45\textwidth]{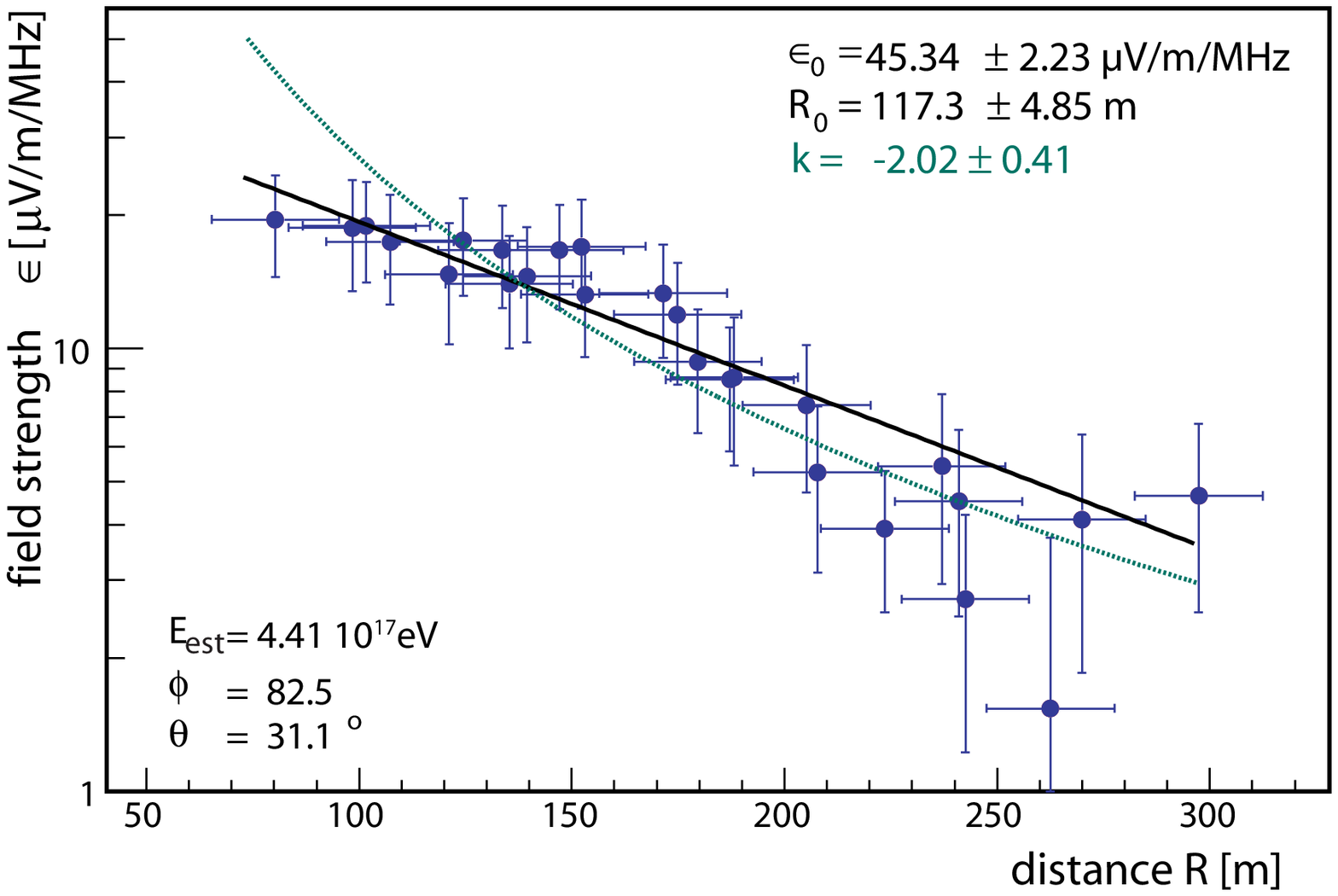}
  \caption{Examples of the radio lateral distribution of individual events measured with LOPES~30. The left example shows a typical event which can be reproduced well with an exponential function. The right example shows an event flattening towards the shower axis. Power-law parameterizations with a spectral index $k$ describe the data close to the shower axis significantly worse than an exponential function.}
  \label{fig:lopes30lateral}
\end{figure*}

\begin{figure*}[htb]
  \centering
  \includegraphics[width=.48\textwidth,clip=true]{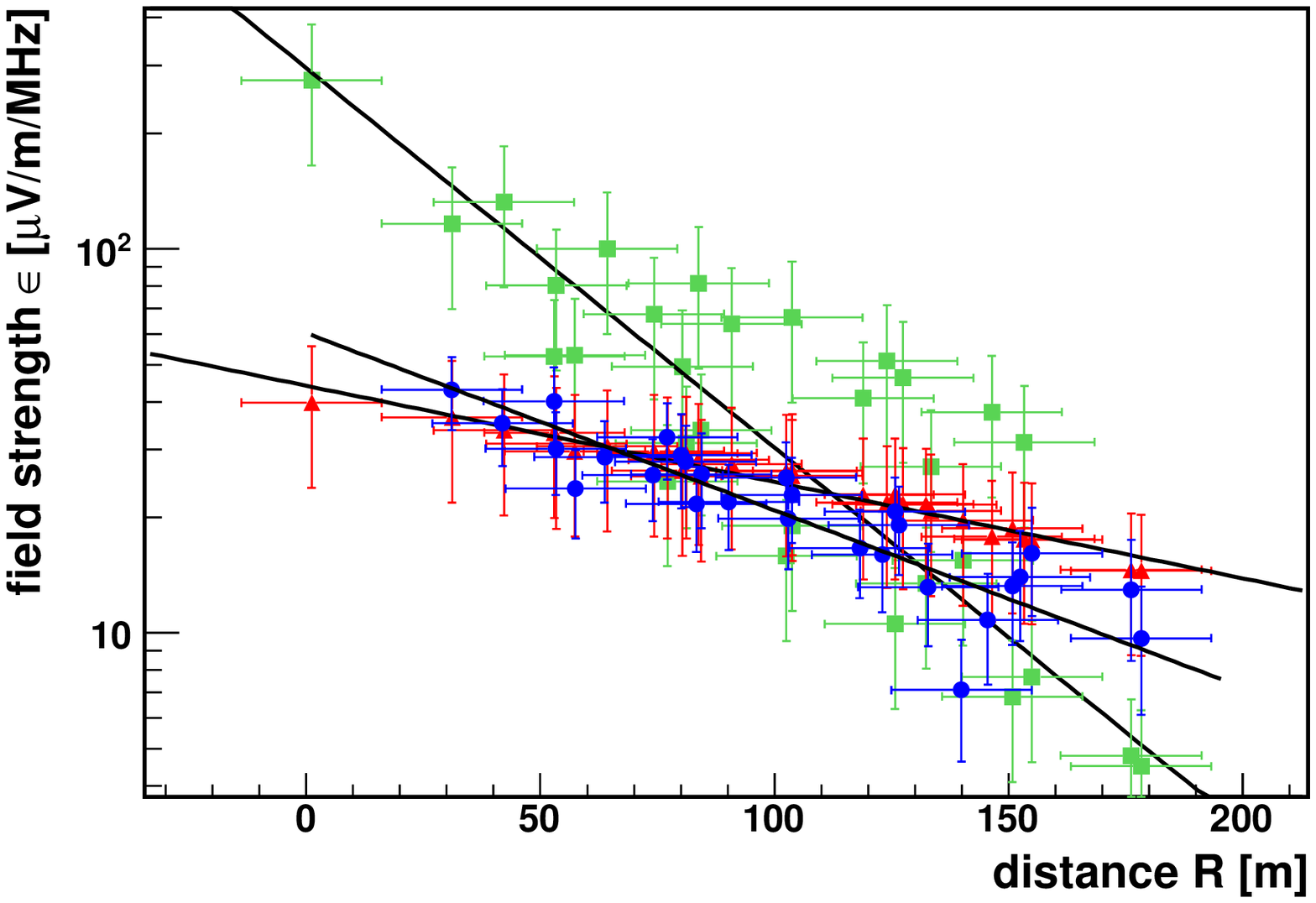}
  \includegraphics[width=.48\textwidth,clip=true]{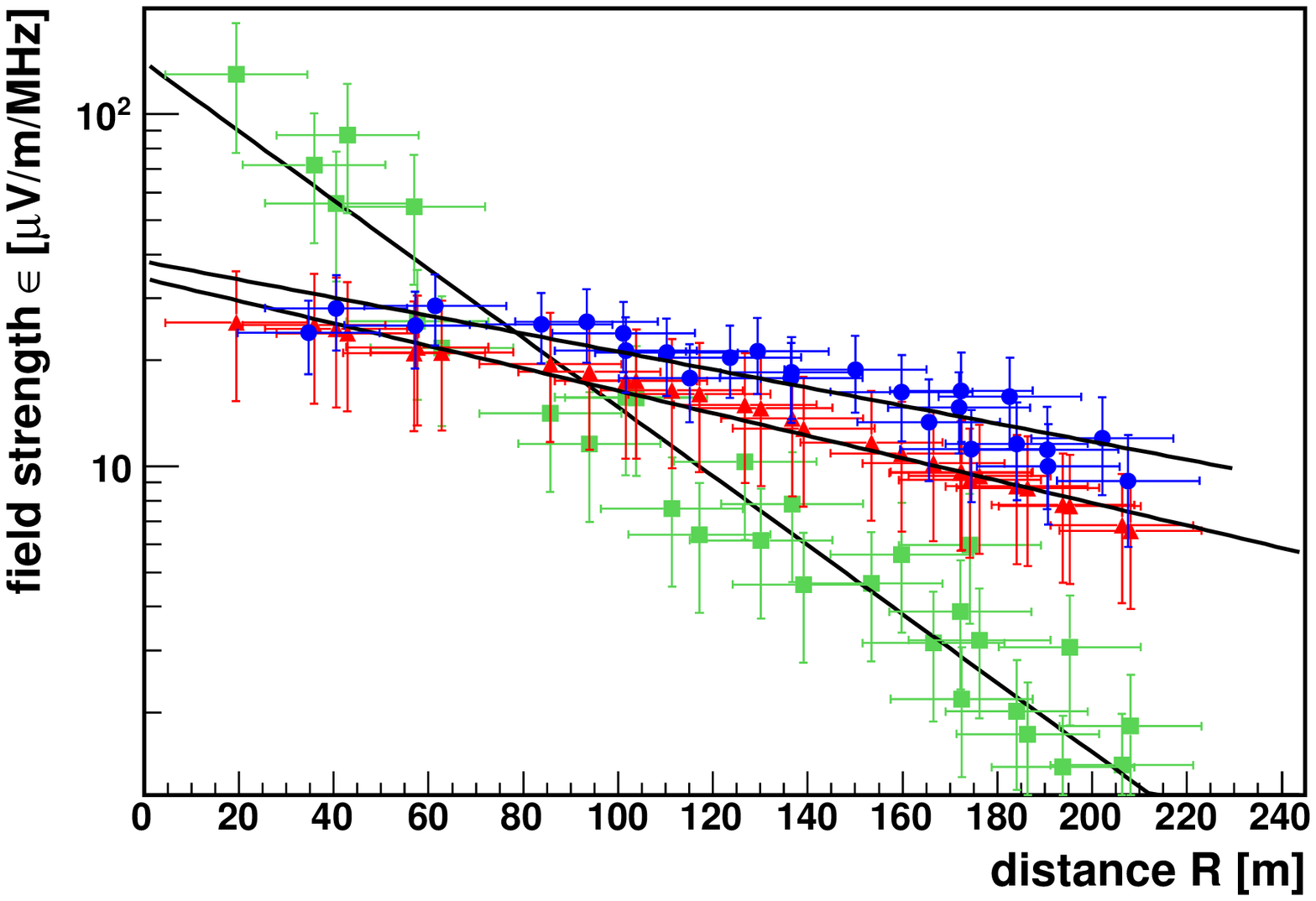}
  \caption{Comparison of two single-event LOPES~30 lateral distribution (blue) with REAS2 (green) and REAS3 (red) simulations of the same events. In the left plot, the REAS3 simulation is for a proton primary, in the right plot for an iron primary. The REAS simulations were based on the KASCADE reconstruction parameters and the simulations of ``typical'' showers with CORSIKA using QGSJET-II.03 and UrQMD1.3.1.}
  \label{fig:lopes30lateralsim}
\end{figure*}

\begin{figure*}[htb]
  \centering
  \includegraphics[width=.48\textwidth]{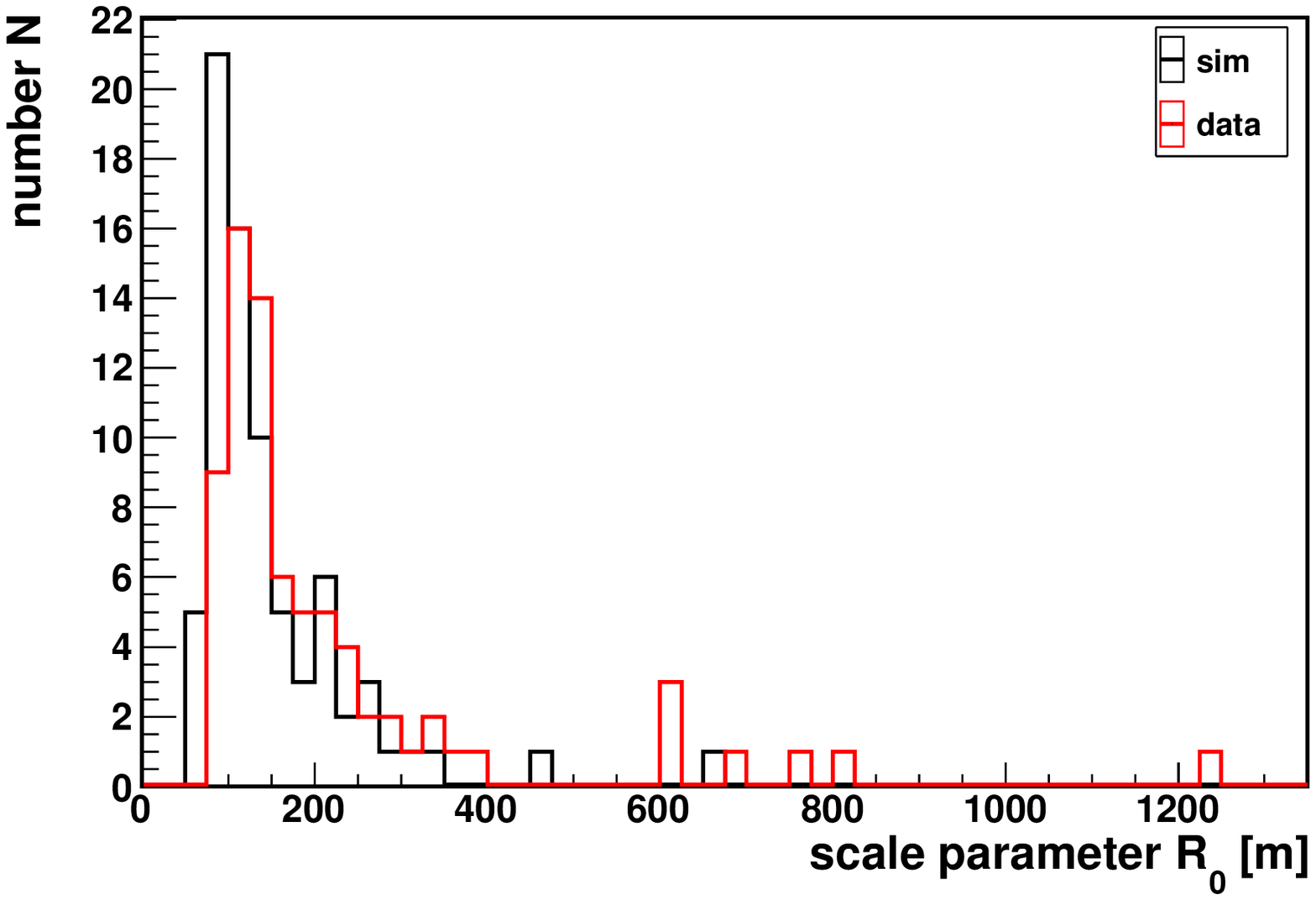}
  \includegraphics[width=.48\textwidth]{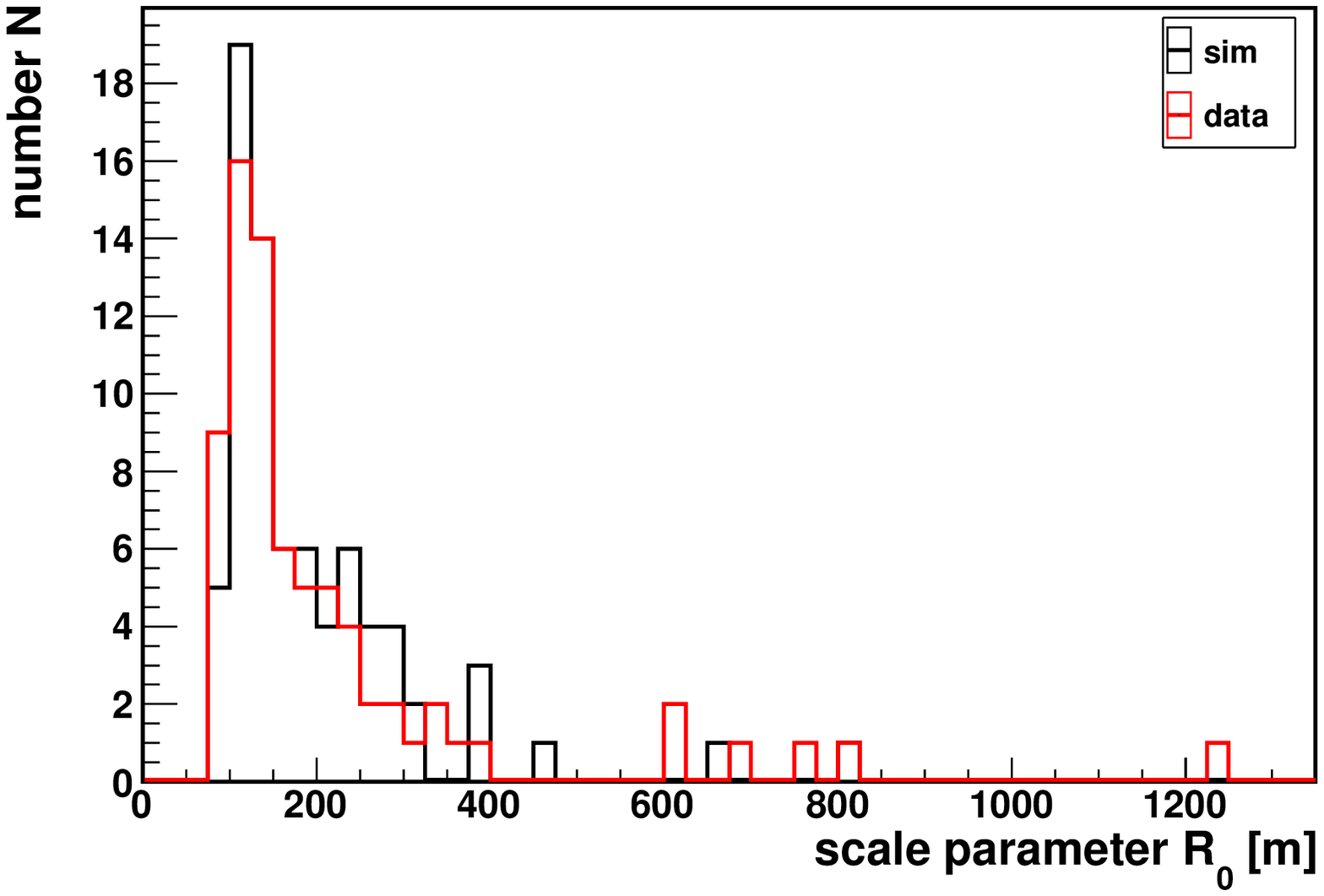}
  \caption{Histograms of the $R_{0}$ distributions for a selection of LOPES~30 events in comparison with those of per-event REAS3 simulations for protons (left) and iron nuclei (right) as primary particles. The REAS3 simulations were performed on the basis of CORSIKA simulations using QGSJET-II.03 and UrQMD 1.3.1. The typical energy of the events is $\approx 10^{17}$~eV.}
  \label{fig:lopes30reashistos}
\end{figure*}

A particular highlight of LOPES~30 results are the detailed, absolute calibrated per-event lateral distributions presented in \citep{ApelArteagaAsch2010}, two of which are shown in Fig.\ \ref{fig:lopes30lateral}. It was found that $\approx 80$\% of events can be well-described with an exponential lateral distribution function with a typical exponential slope of $R_{0}\approx 125$~m. Up to $\approx 20$\% of the events seem to show either completely flat lateral distributions or lateral distributions flattening towards the shower axis. The latter events occur predominantly at high zenith angles and for small lateral distances. In addition, it could be shown that a power-law parameterization is not able to describe the radio lateral distribution near the shower axis as well as an exponential distribution.
These high-quality per-event lateral distributions are ideally suited for comparisons with radio emission simulations such as those performed with REAS2 \citep{HuegeUlrichEngel2007a} and REAS3 \citep{LudwigHuege2010,LudwigHuegeARENA2010}. The agreement in particular with REAS3 simulations is remarkable with respect to both the absolute scale and the lateral slope, as illustrated by the example events shown in Fig.\ \ref{fig:lopes30lateralsim} and the histograms of $R_{0}$ parameters for measurements and REAS3 simulations of both proton- and iron-induced air showers shown in Fig.\ \ref{fig:lopes30reashistos}. An exception are some of the measured events with very flat lateral distributions which cannot always be adequately reproduced by the simulations. Further systematic comparisons taking into account a more sophisticated treatment of detector effects are currently being performed. One effect that can influence such studies is the treatment of noise in the measured data, which has to be carried out with particular care \citep{SchroederARENA2010}.

\begin{figure*}[htb]
  \centering
  \includegraphics[width=.48\textwidth]{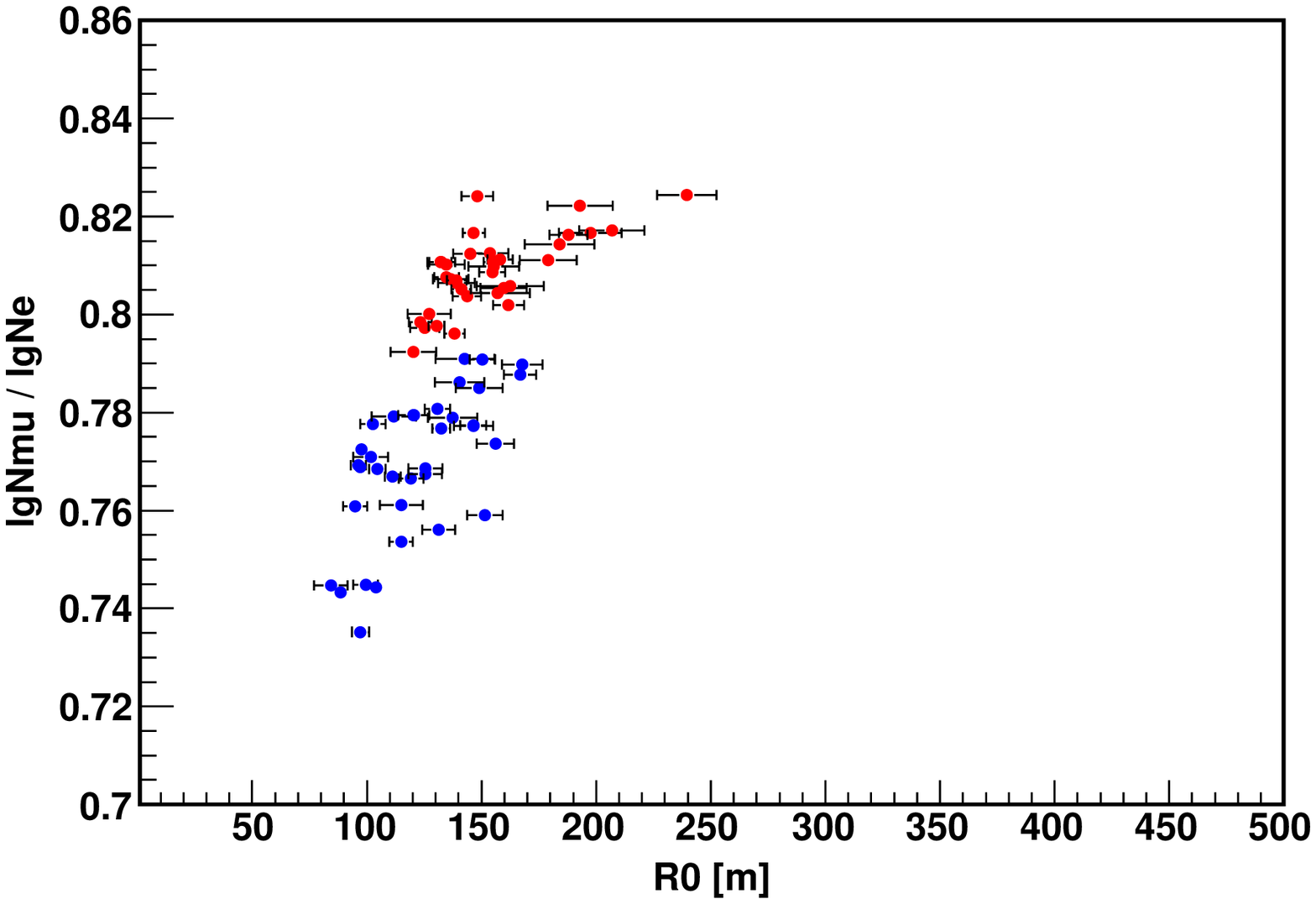}
  \includegraphics[width=.48\textwidth]{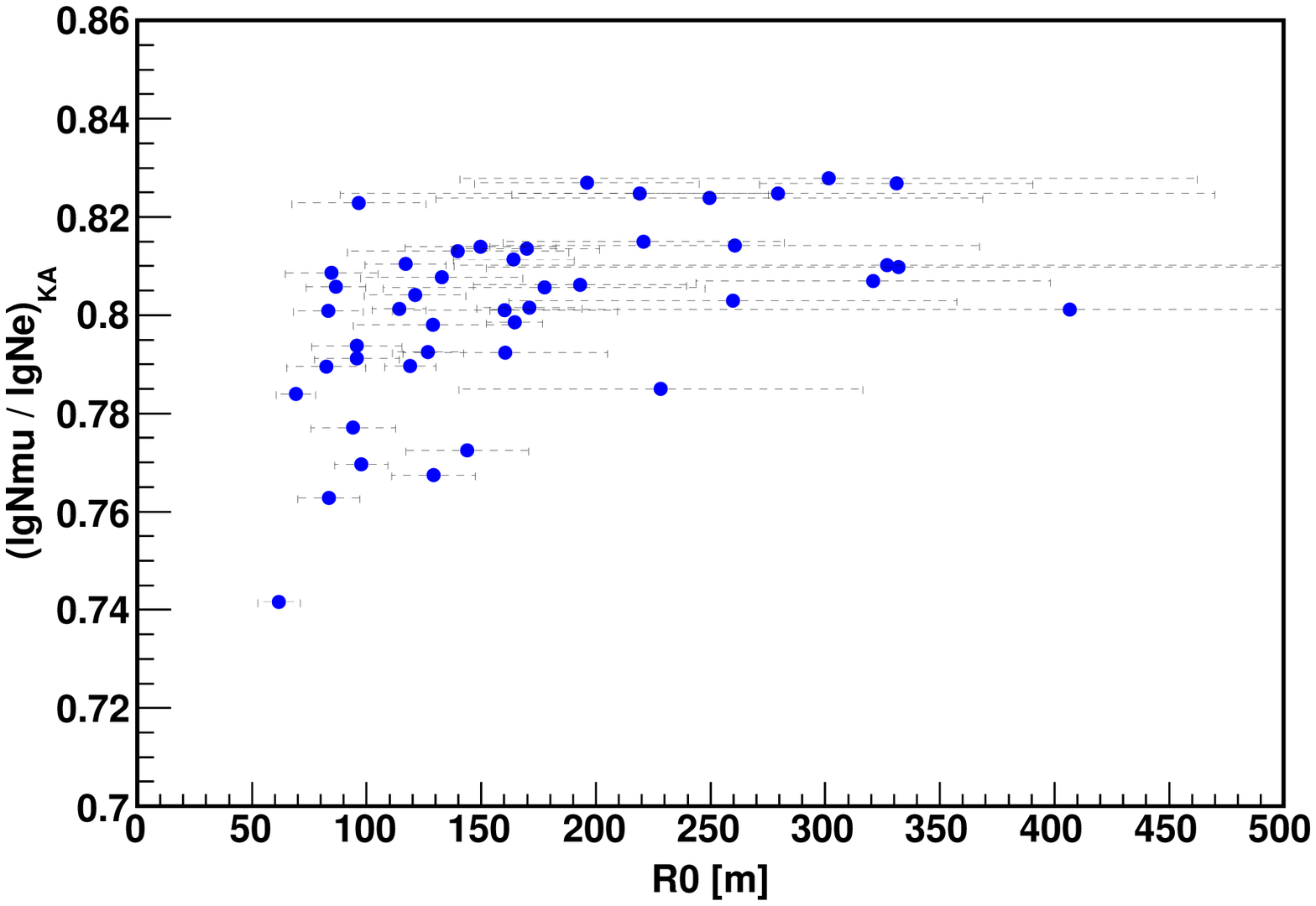}
  \caption{Correlation of the radio lateral slope parameter $R_{0}$ with the ratio of muons and electrons measured by KASCADE for REAS3 simulations (left) and LOPES data (right). In the simulations plot, red points denote iron-initiated showers, blue points denote proton-initiated showers. The REAS3 simulations were performed on the basis of CORSIKA simulations using QGSJET-II.03 and UrQMD 1.3.1. The typical energy of the events is $\approx 10^{17}$~eV.}
  \label{fig:lopescomposition}
\end{figure*}

Another important question being studied with LOPES data is the search for a composition sensitivity of the radio emission. Such a sensitivity in various parameters accessible through radio emission measurements has been predicted by simulation studies \citep{HuegeUlrichEngel2008,LafebreFalckeHoerandel2010}. According to \citep{HuegeUlrichEngel2008}, the steepness of the radio lateral distribution, in case of an exponential fit thus the $R_{0}$ slope parameter, should be linked to the $X_{\mathrm{max}}$ of the air shower and hence the mass of the primary particle. KASCADE-Grande provides us with a reliable mass estimator given by the ratio of muon to electron numbers measured for each individual shower. The fact that air showers initiated by heavier nuclei exhibit a higher fraction of muons can be used in statistical analyses to decompose the air shower flux into different elemental groups \citep{ApelArteagaBadea2009}; an identification on the basis of individual events is, however, next-to-impossible due to shower-to-shower fluctuations.
Preliminary investigations of a possible correlation of the radio lateral slope parameter $R_{0}$ with the muon-to-electron-number ratio provided by KASCADE look promising (cf.\ Fig.\ \ref{fig:lopescomposition}). A very detailed analysis of systematic effects will, however, be necessary before any definitive statements can be made. Furthermore, the prominence of a signature for radio emission sensitivity on the primary cosmic ray mass depends on the composition of the cosmic rays in the energy range observable with LOPES (i.e., at $>\approx 10^{17}$~eV). If the cosmic ray flux at these energies is indeed iron-dominated, as can be expected from KASCADE-Grande results \citep{ApelArteagaBadea2009}, the signature might be somewhat weak.


\section{LOPES~30~pol}

After the measurements with 30 east-west-polarized antennas, LOPES was reconfigured to measure both the north-south and east-west polarization components of the radio emission from extensive air showers. The motivation for this reconfiguration was the fact that polarization studies are an ideal tool to distinguish different radio emission mechanisms. LOPES~30~pol consisted of 15 channels with east-west polarization and 15 channels with north-south polarization. At 5 antenna positions, both polarizations were combined at the same location. An overview of the LOPES~30~pol layout is shown in Fig.\ \ref{fig:lopeslayout}. After a TV transmitter in the LOPES frequency window which had been used as a reference to correct timing delays had been shut down, the beacon transmitter mentioned already in section \ref{sec:analysisprocedure} was set up to perform the phase correction with significantly improved accuracy \citep{SchroederAschBaehren2010}.

\begin{figure}[htb]
  \centering
  \includegraphics[width=.45\textwidth]{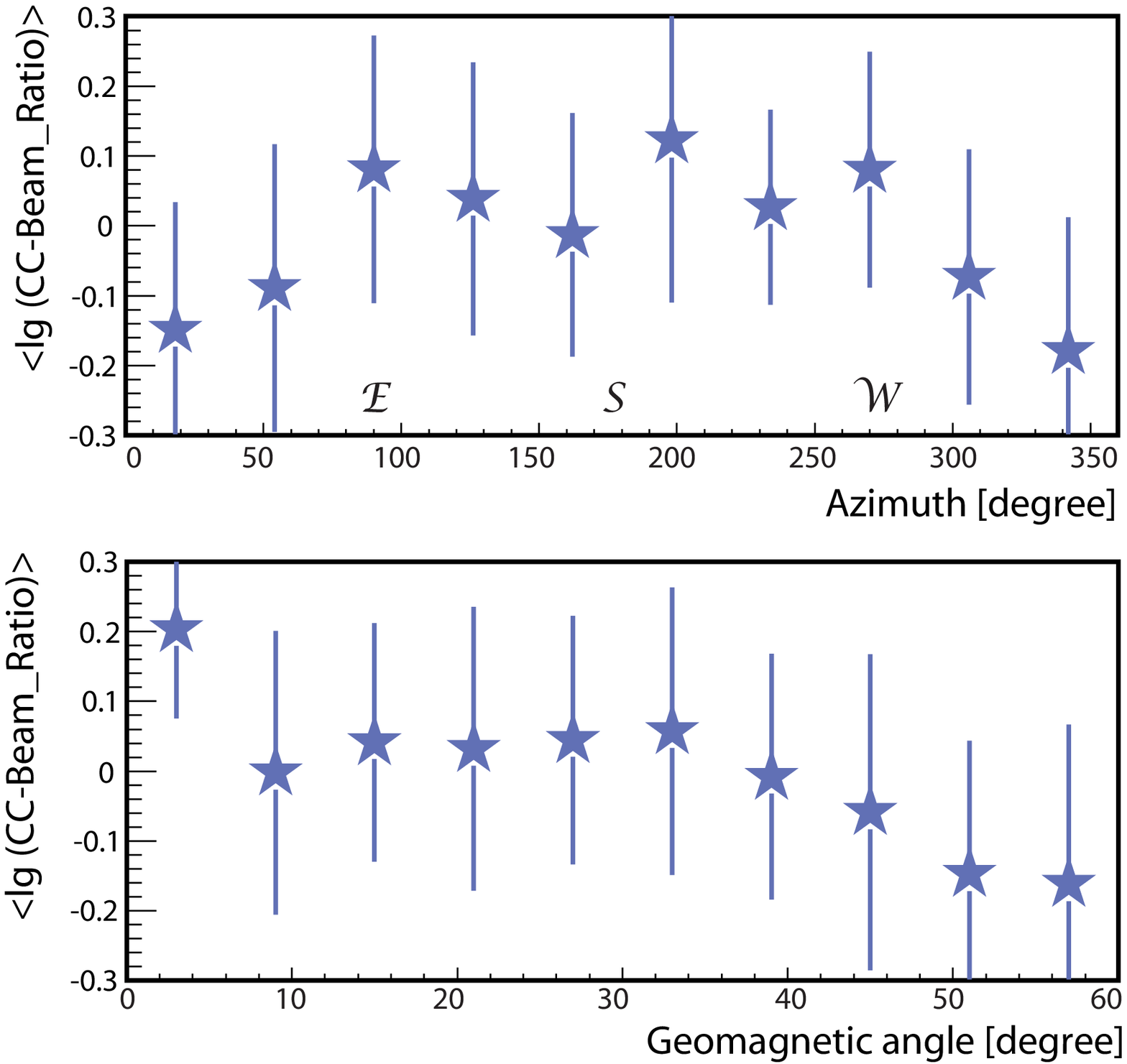}
  \caption{Measurements with the LOPES~30~pol setup show a distinct dependence of the relative strength of the north-south and east-west polarization components (here plotted as ratios of the corresponding CC-beam amplitudes) with both the azimuth angle of the air shower arrival direction (top) and the geomagnetic angle (bottom).}
  \label{fig:lopespol}
\end{figure}

First investigations of the data acquired with the LOPES~30~pol setup have confirmed that the polarization of the radio emission depends on the azimuth angle of the air shower arrival direction, see Fig.\ \ref{fig:lopespol} and \citep{IsarICRC2009}. Further studies are being carried out to analyze in detail the polarization characteristics of the radio emission and their relation to simplified $\vec{v} \times \vec{B}$ models or full-fledged simulations (see, e.g., \citep{SaftoiuARENA2010}).

\section{LOPES~3D}

\begin{figure}[htb]
  \centering
  \includegraphics[width=.3\textwidth]{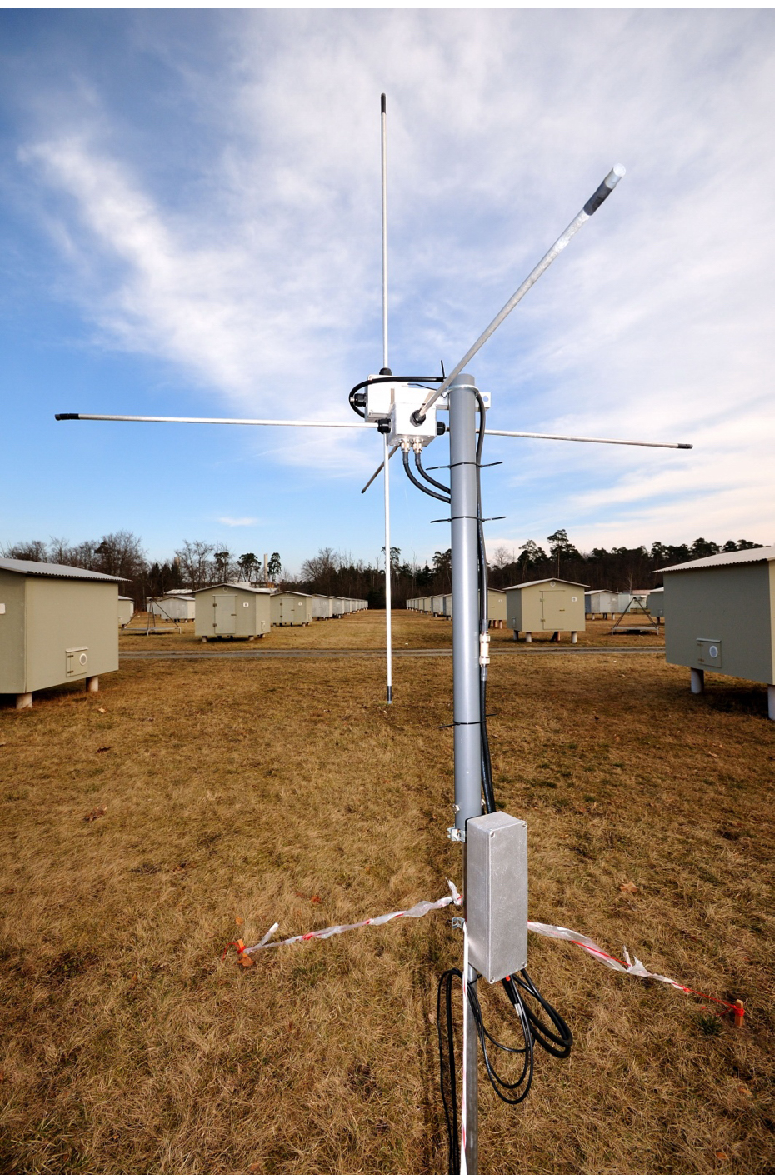}
  \caption{A tripole of the LOPES~3D setup.}
  \label{fig:lopes3Dtripole}
\end{figure}

\begin{figure}[h!tb]
  \centering
  \includegraphics[width=.45\textwidth]{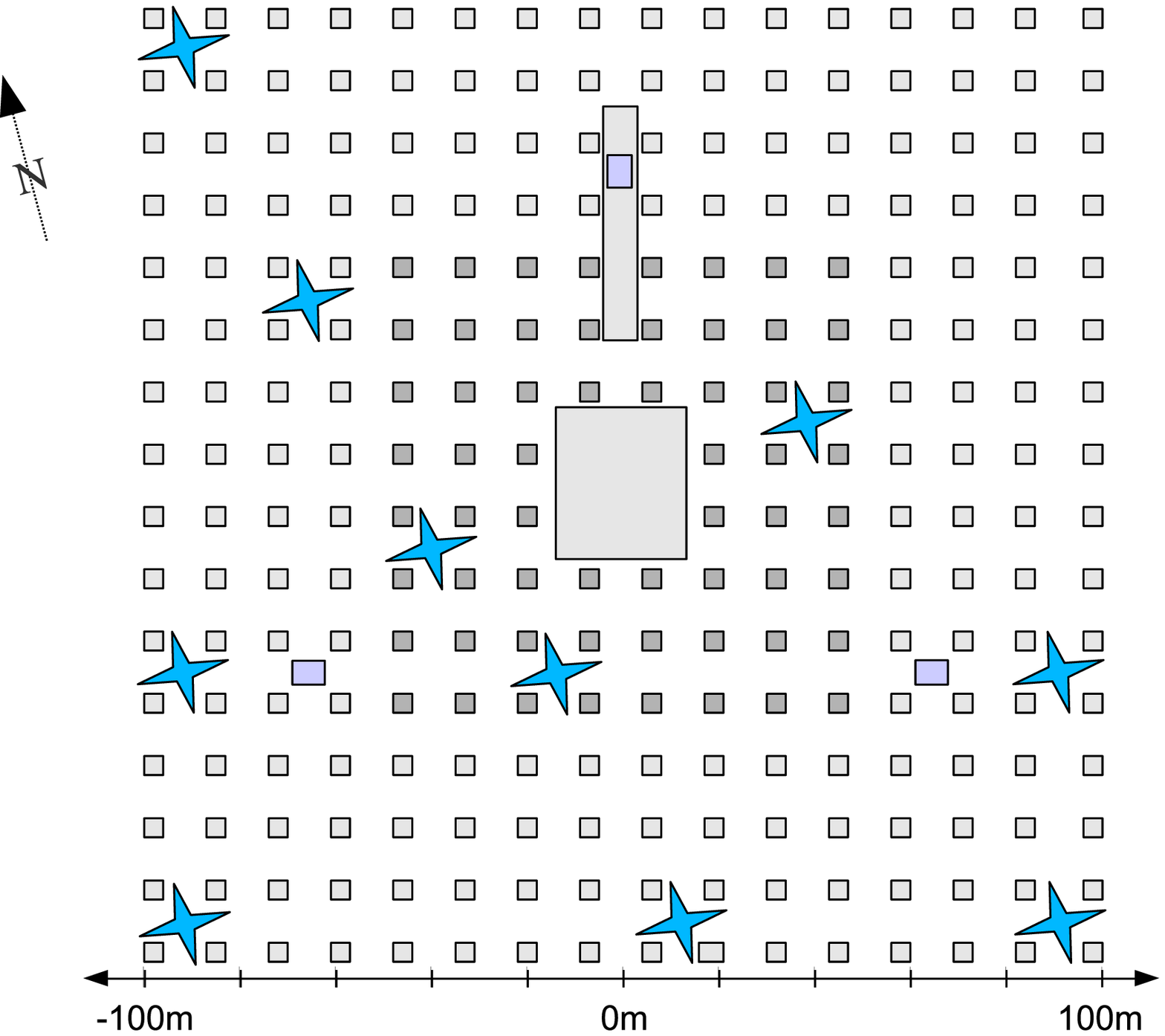}
  \caption{Layout of the LOPES~3D setup consisting of 10 tripoles marked with blue stars.}
  \label{fig:lopes3Dlayout}
\end{figure}

In spring 2010, LOPES was once more reconfigured to the LOPES~3D setup. As the name implies, this setup is targeted at measuring the complete three-dimensional electric field vector instead of just a two-dimensional projection of the electric field on the horizontal plane. This has a number of benefits:
\begin{itemize}
\item{In contrast to a two-dimensional measurement, the complete radio signal is sampled, which leads to an increased signal-to-noise ratio for events with a significant fraction of vertical polarization, in particular highly inclined air showers.}
\item{As electromagnetic waves in the atmosphere are transverse waves, the determination of the complete electric field vector provides information on the propagation direction of the emission already using a single antenna. This information can be used for cross-checks with the arrival direction determined by the arrival time distribution.}
\item{The complete information provided by the three-dimensional electric field vector can potentially help in the development of a polarization-based veto for an advanced radio self-trigger.}
\item{Emission models make predictions on the fraction of vertical polarization of the radio emission. In case of a simplified $\vec{v} \times \vec{B}$ model, e.g., the vertical component should be proportional to the north-south component. With three-dimensional measurements, such predictions can be tested.}
\end{itemize}
For the three-dimensional measurements, the antennas used in LOPES had to be replaced. The electronics, with the exception of the low-noise amplifiers connected to the antennas, remained unchanged. Initial measurements with a SALLA antenna \citep{KroemerSALLA2009} plus a vertical dipole and a self-developed tripole (see Fig.\ \ref{fig:lopes3Dtripole}) consisting of three crossed dipoles have both been successful. For the final setup, the tripole antennas have been selected because they see a clearer signal of the reference beacon and because they represent a more homogeneous design than a SALLA plus a vertical dipole, which helps to minimize systematic uncertainties related to the simulation of the antenna characteristics.
Reconfiguration of LOPES to LOPES~3D began in February 2010. All calibration steps (determination of antenna positions, timing calibration, determination of reference phases, absolute amplitude calibration, simulation of antenna gain patterns, etc.) have been successfully performed. Since May 2010, the LOPES~3D setup with a layout of 10 tripoles as depicted in Fig.\ \ref{fig:lopes3Dlayout} is in stable data taking.


\section{LOPES$^{\mathrm{STAR}}$}

LOPES constitutes an ideal environment for research and development aimed at the large-scale application of the radio detection technique envisaged for example with the Auger Engineering Radio Array (AERA) \citep{HuegePisa2009,FliescherARENA2010}. Developments of new antenna types for radio detection of cosmic rays such as the SALLA \citep{KroemerSALLA2009} and a sophisticated digital self-trigger \citep{SchmidtGemmekeApel2009} have been carried out under the roof of LOPES with the LOPES$^{\mathrm{STAR}}$ setup consisting of 10 dual-polarized antennas distributed within the KASCADE-Grande array (see Fig.\ \ref{fig:lopeslayout}).





\section{Conclusions and Outlook}

The LOPES experiment has been very successful in studying the physics of radio emission from cosmic ray air showers with a variety of setups tailored to specific scientific questions. It has made key contributions to the understanding of the radio emission physics and confirmed theoretical expectations such as the predominantly geomagnetic origin of the emission, the coherence of the radiation in the frequency band up to 80~MHz, the high angular resolution achievable with the radio technique and the reliability of radio measurements in all but the most extreme weather conditions. Furthermore, the absolute calibration of LOPES facilitates quantitative comparisons with the predictions of modern radio emission models, and indeed modern models such as REAS3 are able to reproduce many of the features of the radio emission, including its absolute strength.

Today, cosmic ray radio detection is in a phase of transition from first-generation, small-scale experiments to a full-fledged, stand-alone detection technique for cosmic ray air showers. Major challenges involved with this transition are a reliable and efficient self-triggering of radio signals, a precise determination of the primary energy from radio measurements, and the verification of the predicted composition sensitivity. The future results of LOPES will contribute to solving these challenges and continue to drive the development of the radio detection technique towards its application on large scales for the study of high- and ultra-high-energy cosmic rays.

\section*{Acknowledgements}

Part of this research has been supported by grant number VH-NG-413 of the Helmholtz Association. LOPES and KASCADE-Grande have been supported by the German Federal Ministry of Education and Research. KASCADE-Grande is partly supported by the MIUR and INAF of Italy, the Polish Ministry of Science and Higher Education and by the Romanian Authority for Scientific Research CNCSIS-UEFISCSU (grant IDEI 1442/2008).




\end{document}